\documentstyle[11pt]{article}
\input psfig.sty
\def\singlespace{\baselineskip=13.5pt\lineskip=0pt
\lineskiplimit=-5pt}

\def\title#1{\relax\vspace*{2cm}{\large{\bf #1}}\par\vspace*{13.5pt}}
\def\author#1{{#1}\par\vspace*{13.5pt}}

\def\abstract{\vspace*{27pt}ABSTRACT\par\relax}
\def\section#1{\par{#1}\par}
\def\subsection#1{\par\underline{#1}\par}
\def\subsubsection#1{\par\underline{#1.}\ \ }

\newenvironment{references}{\section{REFERENCES}\vspace*{.5cm}%
\parindent=0pt\frenchspacing%
\parskip=1pt plus 1pt minus 1pt%
\interlinepenalty=1000\tolerance=400%
\pretolerance=10000\hyphenpenalty=10000%
\everypar={\hangindent=1.6pc}
}{}

\newcommand {\et}{\sl et al \rm}

\newcommand {\arcsec}{\mbox{$^{\prime\prime}$}}

\newcommand{\vol}[2]{\bf #1, \rm #2}

\newcommand{\apj}{\sl Astrophys.J. \rm}
\newcommand{\apjl}{\sl Astrophys.J.Lett \rm} 
\newcommand{\mn}{\sl Mon.Not.R.astr.Soc., \rm}
\newcommand{\AnA}{\sl Astron.Astrophys., \rm}
\newcommand{\pasp}{\sl Publ.astron.Soc.Pacif., \rm}

\newcommand{\aj}{\sl Astron.J. \rm}
\newcommand{\anrev}{\sl Ann.Rev.Astron.Astrophys., \rm}
\newcommand{\nat}{\sl Nature, \rm}

\begin{document}
\singlespace
%\doublespace

%%%%%%%%%%%%%%%%%%%%%%%%%%%%%%%%%%%%%%%%%%%%%%%%%%%%%%%%%%%%%%%%%%%%%%%%%%%
%                TITLE     
%%%%%%%%%%%%%%%%%%%%%%%%%%%%%%%%%%%%%%%%%%%%%%%%%%%%%%%%%%%%%%%%%%%%%%%%%%%

\Large {\bf THE AGN/NORMAL GALAXY CONNECTION : SUMMARY} \footnote{Invited review
presented in workshop session of the 32nd COSPAR Meeting,Nagoya, Japan 1998 \\
{\it ``The AGN/Normal Galaxy Connection''}, \\ eds. H. R.
Schmitt, A. L. Kinney and L. C. Ho. \\ To be published in Advances in Space
Research (Oxford: Elsevier) 1999.}

\vskip 1cm  

A. Lawrence
\large\vskip 0.5cm
{\it Institute for Astronomy, University of Edinburgh, \\
Royal Observatory, Blackford Hill, Edinburgh EH9 3HJ, Scotland, UK \\
Email : A.Lawrence@roe.ac.uk \hskip 1.0cm   Tel :  +44-131-668-8356 }

%%%%%%%%%%%%%%%%%%%%%%%%%%%%%%%%%%%%%%%%%%%%%%%%%%%%%%%%%%%%%%%%%%%%%%%%%%%
%               ABSTRACT   
%%%%%%%%%%%%%%%%%%%%%%%%%%%%%%%%%%%%%%%%%%%%%%%%%%%%%%%%%%%%%%%%%%%%%%%%%%%

\vskip 0.5cm {\bf ABSTRACT}
\normalsize

The connection between normal and active galaxies is reviewed, by summarizing
our progress on answering nine key questions. (1) Do all galaxies contain
massive dark objects (MDOs)? (2) Are these MDOs actually supermassive black
holes? (3) Why are the dark objects so dark? (4) Do all galaxies contain an
Active Galactic Nucleus (AGN)? (5) Are the \lq\lq dwarf AGN \rq\rq\ really
AGN? (6) Does AGN activity correlate with host galaxy properties? (7) How are
AGN fuelled? (8) Is AGN activity related to starburst activity? (9) How do
quasars relate to galaxy formation?

%%%%%%%%%%%%%%%%%%%%%%%%%%%%%%%%%%%%%%%%%%%%%%%%%%%%%%%%%%%%%%%%%%%%%%%%%%%
%               BODY OF PAPER 
%%%%%%%%%%%%%%%%%%%%%%%%%%%%%%%%%%%%%%%%%%%%%%%%%%%%%%%%%%%%%%%%%%%%%%%%%%%

\vskip 0.3cm
\section{INTRODUCTION}

Twenty years ago Active Galactic Nuclei (AGN) seemed spectacular but rare
objects, a kind of sideshow compared to the main astrophysical concerns of the
geometry of the Universe, the formation of stars, and the origin of galaxies.
Since then, quasars have become a standard cosmological tool; we have
realised the close connections between the theoretical problems of star
formation and the formation and fuelling of AGN; we have gathered strong
evidence that every galaxy contains a weak AGN or a quiescent black hole;
and there is a growing realisation that the nuclear activity and star
formation histories of the Universe are closely linked. This COSPAR workshop
has brought together a fascinating mixture of scientists to address these
issues. The progress we have made and the distance still to go can be
summarized by looking at nine key questions. In this review I will skim briefly
over the surface of these nine questions.  

\section{DO ALL GALAXIES CONTAIN MASSIVE DARK DARK OBJECTS? }

Although some of the most impressive cases for Massive Dark Objects (MDOs) have
come from gas dynamics and masers (e.g. Marconi \et\ 1997; Miyoshi \et\ 1995)
the most systematic searches have been the stellar kinematics studies. The
review by Kormendy and Richstone (1995) found MDOs in 20\% of E-Sb galaxies
searched, and claimed a correlation between the mass of the dark object and the
stellar mass of the bulge. However, to pass the rigorous filter
of these workers (in particular excluding velocity anisotropy) conditions had to
be favourable - MDOs are easier to confirm in edge-on strongly rotating bulges.
The recent study by Magorrian \et\ (1998) of a large number of galaxies uses
simplified modelling, justified by the results of the more
rigorous studies. These authors find that almost all the galaxies they examine
show clear evidence for central dark objects, and with a correlation between the
mass of the object and the mass of the galactic bulge. Their average ratio
($M_{MDO}/M_{bulge} = 0.006$ ) is moreover twice that suggested by Kormendy and
Richstone (1980). 

These are very exciting results but there are some caveats and worries.
First, the evidence for ubiquitousness comes almost entirely from the most
massive early type galaxies, and so does not yet really tell us that all
galaxies have an MDO, or that the mass correlation is really with bulge mass
rather than total galaxy mass. Second, although the new survey has improved the
MDO--bulge correlation, there is still a very large scatter in MDO size (two
orders of magnitude) at any one galaxy size. Third, these are very {\em large}
black holes compared to the sizes usually invoked in models of quasars and
Seyfert galaxies. 

There seems to be a genuine \lq\lq relic problem\rq\rq. Several authors have
noted that the implied mass density in black holes is an order of magnitude
larger than that expected from the integrated quasar light, and an assumed
accretion efficiency of 10\% (Phinney 1997; Haehnelt, Natarajan and Rees 1998;
Faber, these proceedings). There are various possible explanations. Accretion
efficiency may be much lower than we have assumed; the black holes may grow most
of their mass in some early pre-quasar phase (Haehnelt, Natarajan and Rees
1998); or of course the MDO mass estimates may be wrong. There is
however another very attractive possibility - that there exists a population of
{\em obscured quasars} which outnumber normal quasars by a factor of several.
Direct estimates of the number of obscured AGN are uncertain, depending
sensitively on selection method (Lawrence 1991), but there is strong indirect
evidence - current models of the X-ray background require obscured AGN to
outnumber naked AGN by roughly 3 to 1 (e.g. Comastri \et\ 1995), and these
numbers still do not include objects with columns in excess of a few times
$10^{24}$ which will not contribute significantly to the X-ray background. 

\section{ARE THE MASSIVE DARK OBJECTS ACTUALLY SUPERMASSIVE BLACK HOLES?}

The central massive objects are certainly {\em dark} - in most cases we can
say that the mass-to-light ratio is at least several tens and sometimes hundreds,
thus clearly ruling out normal stellar populations, but leaving other exotic
possibilities such as a cluster of dark stellar remnants. The two most
impressive cases are NGC~4258 and the centre of our own Galaxy, which are
constrained on impressively small size scales and have large minimum central
densities. In the case of NGC~4258, the motions of water masers are detected at
radio wavelengths on milli-arcsec scales, implying a mass density of at least
$4 \times 10^9 M_\odot {\rm pc}^{-3}$ on a scale $r<0.13 {\rm pc}$. Maoz
(1995) argues further that the central object cannot be significantly
extended without measurably distorting the perfect Keplerian rotation curve,
implying a mass density of at least $5 \times 10^{12} M_\odot {\rm pc}^{-3}$ on a
scale $r<0.012 {\rm pc}$. In our own Galaxy, the latest central mass estimates
come from stunning observations of the {\em proper motion} of stars in the
nuclear star-cluster which are consistent with movement around the radio source
Sgr A*  at speeds up to 1700 km s$^{-1}$. This has been made possible by
shift-and-add image sharpening in the near-IR, first on the NTT (Eckart
and Genzel 1996, 1997; Genzel \et\ 1997) and more recently on Keck (Ghez \et\
1998; Morris, these proceedings). The new Keck data have 0.05\arcsec\
resolution and 0.002\arcsec\ positional accuracy; if monitoring is continued we
can even expect to detect {\em accelerations} of the stars (Morris, these
proceeedings). The central dark object in the Galactic Centre has mass $2\times
10^6 M_\odot$ within a radius of 0.01 pc, implying a minimum mass density of
$10^{12} M_\odot {\rm pc}^{-3}$ (Ghez \et\ 1998). Genzel \et\ (1997) have pointed
out that the fact that SgrA* has no measurable proper motion in the (Galactic)
radio frame strongly supports its identification as the location of the black
hole, and indeed on statistical virial grounds this argues that its mass is at
least $10^5 M_\odot$, so that SgrA* probably contains most of the mass causing
the stellar motions.

At the very high minimum densities deduced in NGC~4258 and the Galactic
Centre, any cluster of stellar-size dark objects will have a two-body relaxation
time less than $10^8$ years, so that, following arguments along the lines of
Begelman and Rees (1978), collapse to a black hole seems inevitable (e.g. Genzel
\et\ 1997, Ghez \et\ 1998). We have reached a stage where from an
{\em astronomer's} point of view, the circumlocution \lq\lq massive dark
object\rq\rq\  seems unnecessarily cautious. However from a {\em physicist's}
point of view this hardly seems proof of the existence of supermassive black
holes. How close are we getting to the relativistic regime~? The directly
measured size scales in NGC~4258 and the Galactic Centre are at roughly $10^4$
times the Schwarzschild radius in those systems. If we accept the argument of
Maoz (1995) that the dark object in NGC~4258 can't be distributed and is no
larger than 0.01pc, we are still a factor of a thousand from the event horizon.
If we make the assumption that  the radio source SgrA* must be larger than the
mass causing the stellar motions in the Galactic Centre, then we have reached 15
Schwarzschild radii - but of course this is not a safe assumption at all.
If we accept the virial argument that SgrA* itself is at least  $10^5 M_\odot$,
then the radio source covers 300 Schwarzschild radii (Genzel \et\ 1997). But of
course the virial argument is statistical (we might have been unlucky), and in a
distributed model, there is no generic reason why the radio source can't be
smaller than the whole object. Probably the best evidence that we are
actually dealing with black holes comes from the broad X-ray iron lines in
active objects (e.g Tanaka \et\ 1995), where we seem to be seeing the
signatures we would expect from rotation within a few Scharzschild radii - very
large velocities, a double peak with blue peak stronger, and the whole profile
shifted to the red by gravitational redshift. Given the quality of the data, we
should say that the evidence is extremely tempting rather than completely
convincing, but hopefully AXAF and XMM will settle this question. 

\section{WHY ARE THE DARK OBJECTS SO DARK?}

Fabian and Canizares (1988) first raised the worry that large black holes in
elliptical galaxies ought to be extremely luminous from accretion of the hot gas
that pervades such objects - but they are not. Likewise, the central sources in
M31 and the Galactic Centre are extremely feeble; but gas is clearly present
so one might expect luminosities many orders of
magnitude larger than those seen (Goldwurm \et\ 1994; Melia 1994).
Meanwhile a parallel problem has arisen with the quiescent states of low mass
X-ray binaries, where the deduced accretion rate from the companion star onto
the disc should produce an X-ray luminosity orders of magnitude larger (e.g.
McLintock \et\ 1995; Lasota 1997). The solution proposed by several authors
(Naryan and Yi 1995; Abramowicz \et\ 1995; Fabian and Rees 1995) is the idea of
the Advection Dominated Accretion Flow (ADAF), which may well be the natural
state of affairs at very low accretion rates. Such flows are predicted to have
very low efficiencies (thus solving the black hole darkness problem) and poor
cooling, leading to electron temperatures of the order 10$^9$K. The expected
spectral energy distribution (SED) has two  peaks, one from free-free in hard
X-rays, and another from thermal synchrotron, together with secondary peaks due
to Compton scattering. The magnetic field is deduced from pressure
equipartition. Quite convincing ADAF models have been published for the
Galactic Centre, for NGC~4258, and for soft X-ray transients (Narayan, Yi, and
Mahadevan 1995; Lasota \et\ 1996; Esin, McLintock and Narayan 1997; see also
review by Narayan 1997). Until recently the available data have not tested the
existence of the predicted GHz peak ; however recent high frequency radio and
sub-mm observations (Hernstein  \et\ 1998; Fabian, these proceedings) show that
the ADAF models overpredict the observations by several orders of magnitude. It
seems very hard for ADAF models to escape this blow.

What other possibilities can explain the darkness problem ? Firstly, perhaps a
significant fraction of the expected energy output could emerge as mechanical
outflow rather than as radiation. This seems after all to be the case in SS433,
where the mechanical luminosity is 1000 times larger than the X-ray luminosity
(Watson \et\ 1986). Secondly, accretion flow need not be steady. The
possible mass supply to SgrA* from mass loss in the nuclear star cluster is on
a scale of one parsec, $10^5$ times the Schwarzschild radius. The dynamical
timescale is of the order of a hundred years, but the flow time is likely to be
much longer. These considerations may give us a reasonable idea of the time
averaged accretion flow onto the outer accretion disc, but may not tell us the
current accretion rate onto the black hole. There is a well known thermal
instability which can lead to effective viscosity, and so accretion rate,
changing by many orders of magnitude between high and low states. This is a
popular explanation for dwarf nova and soft x-ray transient outbursts (e.g.
Mineshige, Kim, and Wheeler 1990; Lasota 1997) and has been invoked to explain
the quasar luminosity function (Siemiginowska and Elvis 1997).  

\section{DO ALL GALAXIES CONTAIN AGN? }

We have known since the early 1980s that nearly all galaxies show nuclear
emission lines, and that a third of all galaxies, and most early Hubble types,
show LINER spectra, hinting at but not proving that some kind of weak
quasar-like activity is extremely common (Heckman 1980). The heroic high S/N
spectral survey of 486 galaxies by Ho Filippenko and Sargent (1997 and
references therein) has strengthened this suspicion, showing that $\sim$ 10\% of
galaxies show weak broad H$\alpha$ lines, and almost half show AGN-like
narrow lines. Meanwhile a very large fraction of elliptical galaxies show weak
compact radio cores (Sadler, Jenkins and Kotanyi 1989; Wrobel and Heeschen 1991;
Sadler, these proceedings.) It is now also becoming clear that a large fraction
of very nearby galaxies contain weak nuclear X-ray sources (Colbert, Lira \et\
these proceedings). It seems that the large galaxies are more likely to contain
AGN candidates - see later section. Given that star formation activity in late
type galaxies could actually {\em mask} very weak quasar-like activity, it is
increasingly tempting to believe that ALL galaxies contain some kind of AGN or
AGN remnant. Of course the worry throughout about such objects (LINERS,
weak radio sources, weak X-ray sources) is - are they {\em really} AGN ?

\section{ARE THE UBIQUITOUS LOW LUMINOSITY AGN CANDIDATES REALLY AGN?}

Some LINERs have broad permitted lines and so are proper quasar analogues - but
what about the very common objects that have only narrow LINER spectra ? Ho
(these proceedings) stressed that if the ratio of \lq\lq LINER 2s\rq\rq\  to
\lq\lq LINER 1s\rq\rq\  is similar to the ratio of Type 2 to Type 1 Seyfert
galaxies, then a large fraction of all LINERs would be explained. The
expectation that LINER 2s can be obscured versions of LINER 1s has been
spectacularly confirmed by the discovery of polarised broad H$\alpha$ in
NGC~1052 (Barth 1998, PhD thesis - diagram shown by Ho in these proceedings),
showing the existence of an obscured BLR revealed in reflection, just as in
NGC~1068. On the other hand, UV spectroscopy by Maoz \et\ (1998) of the compact
UV sources seen in some LINERs shows very clear signatures of winds from hot
young stars, showing that such objects contain young stellar clusters. Maoz \et\
show that those objects with clear stellar signatures are at least an order of
magnitude less luminous in X-rays. It may be
that LINERS are a genuinely heterogeneous class. On the other hand, the
emission from such a young cluster could actually {\em mask} the presence of a
very weak or obscured AGN. 

The X-ray emission from broad-lined LINERs seems quite consistent with other
properties (Koratkar \et\ 1995; Fabbiano 1996; Serlemitsos, Ptak, and
Yaqoob 1996; Terashima \et\ 1997; Terashima, these proceedings) and indeed one
cannot distinguish \lq\lq dwarf AGN\rq\rq\  whose narrow-line components are
LINER-like from those whose narrow-line components are Seyfert-like. However
worries have been raised that seem to distinguish dwarf AGN from more luminous
objects like Seyfert galaxies and quasars. (i) It has been suggested that they
do not vary, or vary less than Seyfert galaxies (Shields and Filippenko 1992;
Ho, Filippenko, and Sargent 1996; Ptak \et\ 1998; Awaki, these proceedings).
However the least luminous known AGN, NGC~4395, has been shown to vary rapidly,
with colour changes just like those seen in Seyfert galaxies (Lira \et\ 1998).
It may well be that NGC~4395, a dwarf galaxy, has a very small black hole,
whereas many other LINERs have large black holes with low accretion rates.
Further careful quantification is needed on the variability question. (ii) A
second worry is that dwarf AGN tend to have no Big Blue Bump, but instead have
steep optical-UV spectra, with $\alpha\sim 2$, and possibly also have a mid-IR
excess compared to quasars (Ho, Filippenko and Sargent 1996; Barth \et\ 1996;
Ho, these proceedings).

A possible explanation is dwarf AGN have very cool \lq\lq bumps\rq\rq\  rather
than absent ones. Empirically, their steep spectra are consistent with the
trends claimed by Kriss (1988), Wandel and Mushotzky (1989) and Zheng and Malkan
(1993). Quasar SEDs systematically steepen from optical to UV, reaching
$\alpha\sim 2$ in the far UV (Zheng \et\ 1997), and in any one spectral range
steepen systematically as luminosity is lowered, from $\alpha=0$ for the most
luminous quasars to $\alpha=2$ for dwarf AGN. Previous attempted explanations
have concentrated on changing bump strength, but an attractive possibility is
that characteristic  temperature changes with luminosity. Lawrence (1998a)
describes how multi-temperature models scale in a characteristic fashion and
produce an excellent fit to the trends of SED shape with luminosity.

\section{DOES AGN ACTIVITY CORRELATE WITH HOST GALAXY PROPERTIES?}

One of the persistent facts about local AGN is that more or less
without exception radio-loud AGN are in elliptical galaxies, whereas radio
quiet AGN are in spirals. There has been a debate about whether such a
distinction continues to hold for the hosts of low-redshift quasars (see eg
Taylor \et\ 1996 and references therein). The most careful study of quasar hosts
so far is being undertaken with HST by Dunlop and collaborators. At this
workshop, Kukula showed evidence that {\em all} the most luminous quasars live
in giant ellipticals, regardless of radio loudness. However residuals from the
smooth $r^{1/4}$ fits often show much disturbed structure suggesting mergers,
complicating the interpretation. It has been suggested that mergers are central
to the process of formation of both elliptical galaxies and quasars (Sanders
\et\ 1988; Kormendy and Sanders 1989).

For many years there has been tantalising but not completely convincing
evidence that quasar luminosity correlates with host galaxy luminosity (e.g.
Lawrence 1993 and references therein). McLeod and Rieke (1995) have argued that
rather than being a simple correlation between those quantities, there is an
{\em upper envelope} to quasar luminosity which is proportional to galaxy size,
and that quasars and Seyferts are consistent with the same relationship. In
other words, big galaxies can have big or small AGN, but small galaxies can only
have small AGN. A possible simple explanation is that black hole mass is on
average proportional to galaxy mass (as local dynamical studies seem to
indicate), but that a given black hole can have any accretion rate below a
maximum given by the Eddington rate. The observed upper envelope is at least
roughly consistent with that expected from the Magorrian \et\ $M_H/M_{bulge}$
relationship (McLeod 1997).

In their study of weak radio cores, Sadler, Jenkins and Kotanyi (1989) made
essentially the same point concerning the wedge-like statistical relation
between AGN power and galaxy luminosity, but went somewhat further, constructing
the bivariate luminosity function, and trying various ways of quantifying the
relationship, such as correlating galaxy luminosity $L_B$ with the 30th
percentile radio power $P_{30}$. (Radio astrononmers were always better at
statistics ..) They found that $P_{30} \propto L_B^{2.2}$. On the other hand it
seems that optical emission line strength is proportional to $L_B$ (Sadler,
these proceedings; Sadler, Jenkins and Kotanyi 1989). Lira \et\ (these
proceedings) have searched for weak nuclear X-ray sources in a volume limited
sample of nearby galaxies. Nearly all the non-detections were in the smaller
galaxies, and once again there appeared the all too familiar wedge-like pattern,
consistent with the idea of an upper envelope to nuclear X-ray luminosity being
proportional to galaxy luminosity. Of course we can't be sure yet whether these
weak X-ray sources are really AGN.

A significant puzzle, noticed by Sadler, Jenkins and Kotanyi, but now
significantly strengthened, is that optical, X-ray, and emission line activity
all seem to correlate roughly linearly with galaxy luminosity, whereas radio
core power is much more sensitive, going as something like $L_B^{2-3}$.
Whatever the explanation, this might help to make sense of the radio loudness
dichotomy if the connection is specifically with spheroid component as often
assumed.

\section{HOW ARE AGN FUELLED? }

A clear analysis of the fuelling problem is given in Shlosman, Begelman, and
Frank (1989), and a useful collection of reviews, results, and theories can be
found in the proceedings of the 1993 Kentucky conference {\it Mass Transfer
Induced Activity in galaxies} (Shlosman 1994), and also in the proceedings
of the 1996 Saas-Fee meeting, {\it Galaxies : Interactions and Induced Star
Formation} (Kennicut, Schweizer, and Barnes 1998). It is an exceedingly difficult
problem. Material has to change radial scale by perhaps nine orders of magnitude
and lose something like five orders of magnitude of specific angular momentum.
(Phinney 1994 makes this point particularly clearly not only in words but with
a wonderful cartoon which I shamelessly stole for my talk at this conference).
There is no shortage of ideas, most of which involve some kind of gravitational
instability or non-axisymmetric potential. The problem becomes not so much 
\lq\lq can it be done ?\rq\rq\  but rather \lq\lq which of these ACTUALLY
happens ?\rq\rq\ . However there will never be a simple theory of AGN fuelling.
Each candidate process manages to shrink material by typically a factor of a few
- so clearly a whole {\em sequence} of processes is needed.  We can crudely
divide the problem into four stages - (1) galaxy scale to central regions; (2)
central regions to ten parsec scale; (2) ten parsec scale to accretion disc; (3)
accretion disc to event horizon.

Most of the observation and argument in recent years has concerned stage
(1), and the role of interactions, mergers, and large-scale bars. It seems a
particularly good bet that such processes are involved in triggering central
starbursts. Most of this work concerns current day activity in existing
galaxies, but some authors argue that the chaotic
dynamics and clump interactions in the process of galaxy formation itself
naturally leads to collapse of some central fraction of the gas, which may be
closely related to the peak of quasar activity (Lake, Katz, and Moore 1998;
Lake, Noguchi, these proceedings). Stage (2), from hundreds of parsecs to a few
parsecs, has received less detailed attention. Theoretical possibilities include
gravitational instabilities in the self-gravitating gas disk formed in stage
(1), possibly as a cascade of bar instabilities (Shlosman, Frank and Begelman
1989); disruption of the disc by star formation; or magnetic
braking (Krolik and Meiksin 1990). Some clues may be coming from gas morphology
in the central regions. CO mapping of galaxies has shown that the central cold
gas can have a variety of morphologies - rings, bars, central peak, twin peaks.
These seem unlikely to be equilibrium dynamical structures and instead may tell
us about the evolution of the central regions. Some important distinctions seem
to be emerging - galaxies with AGN usually have CO rings and small ratios of
gas mass to dynamical mass, whereas galaxies with HII spectra have CO bars and
large ratios of gas mass to dynamical mass (Sakamoto \et\ 1997; Ishizuki, these
proceedings). This is a complicated subject but we may be close to putting
together a feasible history of episodic collapse and star formation.

If Roberto Terlevich is right (e.g. Terlevich \et\ 1992), stages (3) and (4) are
not needed, as the AGN phenomenon is actually an exotic form of starburst on
parsec scales in a high density environment. In the black hole model there is a
long way to go to the event horizon, but somewhere on the parsec scale we will
reach a point where the gravitational field of the hole dominates over the
stellar field, so that the final fate of the material seems inevitable. It is
tempting to believe that once a ten parsec scale gas disk has been produced in
stage (2), it can fuel the black hole slowly and steadily by some local
viscosity mechanism. However as explained by Begelman (1994) and Shlosman,
Begelman and Frank (1989) such a giant accretion disc picture has serious
problems - the inflow time is of the order of 10$^9$ years or more, and the
accretion rate will be too small to power quasar luminosities, unless the
density is large, in which case the disc becomes gravitationally unstable to
local clumping (in which case it will probably form stars, cease to dissipate,
and so stop flowing in). It seems likely that even in this inner
region, large scale (rather than local) gravitational or magnetic effects are
needed to re-distribute angular momentum, which will probably happen in lurches
rather than in a nice steady fashion. An interesting alternative is some kind
of hot accretion flow, which can potentially support a faster flow and more mass
without becoming unstable (Shlosman, Begelman and Frank 1989).

So far I have assumed that material needs first to be assembled from large
distances. An alternative is that material is supplied from a nuclear star
cluster. Early versions of such models explored disruption of stars (e.g. Hills
1975) but more recent work concentrates on stellar mass loss and supernovae,
the fuelling from which will evolve with time following the formation of the
cluster (Norman and Scoville 1988; Murphy, Cohn, and Durisen 1991; Williams and
Perry 1994). Shlosman, Begelman and Frank (1989) argue that the supply rate
from mass loss is unlikely to be enough to power luminous quasars. However if
quasars are short-lived this may not be relevant, as the mass loss serves to
accumulate a reservoir of material which can later be accreted. Strong support
for developing models of this kind comes from the fact there is now good
evidence that such compact nuclear star clusters really exist - for example in
the Galactic Centre (Krabbe \et\ 1995), in NGC~1068 (Thatte \et\ 1997), and in
a variety of other nearby galaxies (Ho 1997). It may be that this a late-stage
phenomenon connected with recurring low-level activity in recent epochs; but it
has also been suggested that the large abundances seen in high redshift quasars
requires a starburst closely associated with quasars in both time and space
(Hamman and Ferland 1992).

Finally we arrive at the accretion disc. This is a much more mature problem
but cannot be considered completely solved. Reconsideration of the role of
advection has recently shaken the subject up and the origin of viscosity is 
still unclear. The most promising choice for viscosity is thought to be the
Balbus and Hawley magnetic instability (Balbus and Hawley 1991) but in this
case the disc will not behave at all like a standard \lq\lq $\alpha$
disc\rq\rq\ . Energy release will be above the disc rather than inside it
(Begelman and De Kool 1991; Begelman 1994). Accretion is not necessarily steady.
Indeed Siemiginowska and Elvis (1997) use predicted variations in accretion rate
due to thermal instability to explain the quasar luminosity function. 

Finally, an important generic point. At every stage of fuelling there is good
reason to expect that activity will be {\em episodic}.  As Krolik and Meiksin
(1990) point out in their discussion of hundred--parsec scale magnetic braking,
conservation of angular momentum demands that whenever some of the material goes
in, other stuff goes out, so that reservoirs tend to evacuate as they \lq\lq dump
down\rq\rq\ to the next stage. This naturally leads to episodes of accretion.
The same basic point applies at all stages. Other feedback loops seem likely
to operate. For example, various authors have suggested that in nuclear star
clusters feedback between accretion, central radiation, winds, mass loss, and
gravitational instability  will lead to episodes of star formation, mass
accumulation, and nuclear activity in turn (e.g. Bailey and Clube 1978; Williams
and Perry 1994; Morris, these proceedings). Accretion discs may be subject to a
thermal limit cycle (see earlier discussion), and accretion near the Eddington
limit may be self-limiting and erratic. Finally,
fuelling may be a stochastic event that occurs when one particular molecular
cloud with low angular momentum intersects the black hole (Sanders 1981). An
understanding of these time-dependent processes may be important for
understanding quasar evolution. 

\section{IS AGN ACTIVITY RELATED TO STAR BURST ACTIVITY?}

There is good circumstantial evidence that vigorous star formation is nearly
always associated with quasar-like activity, and that much of the
long-wavelength continuum energy distribution is actually from the starburst
(Lawrence 1998 and references therein). But is there an actual causal connection
? It may be that starbursts always precede AGN activity, in a grand sequence of
galaxy interaction -- infall -- starburst -- further infall -- quasar (e.g.
Sanders \et\ 1988). On a smaller scale, parsec scale starbursts may always be
closely connected with quasar-like activity with the causal connection going in
both directions (see previous section) and indeed it has been argued that
parsec scale starbursts could be the whole explanation of quasar-like activity
(Terlevich \et\ 1992). Alternatively the large and small scale processes could
be separate phenomena. Perhaps galaxy mergers cause starbursts which collapse
no further, whereas AGN activity and fuelling is entirely connected with small
scale structures formed at early times. This is one reason why we would still
like to answer that fashionable question of the late 80s - are the
ultraluminous IRAS galaxies (which are very frequently mergers) really
starbursts or are they obscured AGN ? Mid-IR spectroscopy from ISO seems to
show clear cases of each, but with most objects being starbursts and a large
minority being AGN (Genzel \et\ 1998; Lutz \et\ 1998). X-ray studies also
suggest a mixture (Rigopoulou \et\ 1995; Nakagawa, these proceedings).
Re-assuringly, mid-IR and X-ray classifications seem to agree reasonably well
(Lutz, these proceedings). As one might have guessed, ULGs are a heterogeneous
class, so one must be careful drawing conclusions from them.

\section{HOW DO QUASARS RELATE TO GALAXY FORMATION?}

The peak of quasar activity at z=2-3 is suspiciously
similar to the predicted epoch of spheroid formation in cosmological theories,
suggesting a close connection between quasars and galaxy formation (e.g.
Efstathiou and Rees 1988, Haehnelt and Rees 1993). Of course one (probably)
needs a galaxy before one can get a quasar; the subsequent decline may be
connected with a decline in fuelling (e.g. Small and Blandford 1992) but this
is not yet clear. Alternatively some authors have suggested that quasar
activity actually has a causal role in triggering or inhibiting galaxy formation
(Chokshi 1997; Silk and Rees 1998). Our
perspective on such questions is changing rapidly. For three decades quasar
evolution was a hard observed fact, whereas galaxy formation was a
theoretical blur. This situation is now changing dramatically as galaxies are
being detected at high redshift, and we can begin to construct the cosmic
history of the star formation rate (Madau \et 1996). It has been
noted that the evolution of quasar luminosity density tracks the cosmic star
formation rate very closely (Dunlop 1997; Boyle and Terlevich 1998). Silk and
Rees (1998) argue that the star formation rate peaks later (z=1-2) than quasar
activity, and suggest a feedback loop between winds from early quasars and
spheroid formation. However, it now seems clear that the high-redshift star
formation rate is higher than had been thought. The optically selected high-z
sources have significant reddening (Pettini \et\ 1997) and the star formation
rate deduced from the new population of faint sub-mm sources is a factor of
several higher, improving the close agreement with the shape of quasar
luminosity density evolution (Hughes \et\ 1998, Blain \et\ 1998). 

The faint sub-mm sources are generally assumed to be starbursts,
the most striking conclusion being that most star formation at early times is
occurring at any one time in a small number of luminous bursts. But could the
faint sub-mm sources be AGN ? Almaini,
Lawrence and Boyle (1998) calculate predicted obscured AGN counts in the sub-mm
(by requiring that the number of obscured AGN matches that required in X-ray
background models) and find that around 5 - 20\% of the detected sources are
probably AGN. This number is sensitive to assumed cosmology and the form of
high-z evolution, such that this fraction is very uncertain, and likely to
increase to even fainter fluxes. Whether the faint sub-mm sources are
starbursts or quasars, they are certainly things going BANG, and such objects
dominate the energetics of the young universe. This is in strong contrast to
today, when the luminosity density in AGN and starbursts combined is a tiny
fraction of the total galaxian luminosity density. History belongs to the
heroes, but the meek shall inherit the Earth.

%\begin{figure}
%\vskip -3.0cm
%\hskip 3.0cm
%\psfig{figure=YOUR_FIGURE.ps,width=10cm,height=12cm}
%\vskip -0.5cm
%Fig. 2. This is the example of a Figure.
%\vskip -1cm
%\end{figure}

\end{document}